\documentclass[aps,jcp,preprint,superscriptaddress]{revtex4}
\usepackage{graphicx}
\usepackage{amsmath}
\usepackage{dcolumn}
\usepackage{bm}

\begin{document}
\title{Cold and ultracold NH--NH collisions: The field-free case}

\author{Liesbeth M.~C.~Janssen}
\affiliation{Radboud University Nijmegen, Institute for Molecules and Materials,
Heyendaalseweg 135, 6525 AJ Nijmegen, The Netherlands}
\author{Piotr S.~\.{Z}uchowski}
\affiliation{Department of Chemistry, Durham University, South Road, Durham, DH1 3LE, 
United Kingdom}
\author{Ad van der Avoird}
\affiliation{Radboud University Nijmegen, Institute for Molecules and Materials,
Heyendaalseweg 135, 6525 AJ Nijmegen, The Netherlands}
\author{Jeremy M.~Hutson}
\email[Electronic mail: ]{J.M.Hutson@durham.ac.uk}
\affiliation{Department of Chemistry, Durham University, South Road, Durham, DH1 3LE, 
United Kingdom}
\author{Gerrit C.~Groenenboom}
\email[Electronic mail: ]{Gerritg@theochem.ru.nl}
\affiliation{Radboud University Nijmegen, Institute for Molecules and Materials,
Heyendaalseweg 135, 6525 AJ Nijmegen, The Netherlands}

\date{\today}

\begin{abstract}
We present elastic and inelastic spin-changing cross sections for cold and
ultracold NH($X\,^3\Sigma^-$) + NH($X\,^3\Sigma^-$) collisions, obtained from
full quantum scattering calculations on an accurate \textit{ab initio} quintet
potential-energy surface. Although we consider only collisions in zero field,
we focus on the cross sections relevant for magnetic trapping experiments.  It
is shown that evaporative cooling of both fermionic $^{14}$NH and bosonic
$^{15}$NH is likely to be successful for hyperfine states that allow for
$s$-wave collisions. The calculated cross sections are very
sensitive to the details of the interaction potential, due to the presence
of (quasi-)bound state resonances.
The remaining inaccuracy of the \textit{ab initio} potential-energy surface 
therefore gives rise to an uncertainty in the numerical cross-section values.
However, based on a sampling of the uncertainty range of the \textit{ab initio}
calculations, 
we conclude that the exact potential is
likely to be such that the elastic-to-inelastic cross-section ratio is
sufficiently large to achieve efficient evaporative cooling.
This likelihood is only weakly dependent on the size of the channel basis set
used in the scattering calculations.

\end{abstract}

\maketitle

\section{Introduction}
Cold ($T < 1$ K) and ultracold ($ T < 1$ mK) molecules offer a wide variety of
applications in condensed-matter physics \cite{micheli:06}, high-precision
spectroscopy \cite{lev:06,bethlem:09,tarbutt:09}, physical chemistry
\cite{meerakker:05, gilijamse:06,gilijamse:07,campbell:08,krems:08}, and
quantum computing \cite{demille:02,andre:06}. In the last few years, techniques
have been developed that either form (ultra)cold molecules by pairing up
pre-cooled atoms, e.g.\ by photoassociation \cite{jones:06} or Feshbach
association \cite{kohler:06}, or by cooling the molecules directly.  Examples
of the latter approach include buffer-gas cooling \cite{weinstein:98} and Stark
deceleration \cite{bethlem:03}.

A promising candidate for direct-cooling experiments is the NH biradical. NH in
its electronic $X\,^3\Sigma^-$ ground state has been cooled from room temperature
using a helium buffer gas and trapped in a magnetic field
\cite{krems:03a,campbell:07,hummon:08,campbell:08}.
Stark deceleration and electrostatic trapping experiments have been performed
on metastable NH($a\,^1\Delta$), which, in contrast to the $X\,^3\Sigma^-$
ground state, exhibits a linear Stark effect.  The decelerated
NH($a\,^1\Delta$) molecules can be converted to the ground state by excitation
of the $A\,^3\Pi \leftarrow a\,^1\Delta$ transition followed by spontaneous
emission \cite{meerakker:01,hoekstra:07}. The resulting NH($X\,^3\Sigma^-$)
molecules may subsequently be accumulated in a magnetic trap.

Direct-cooling techniques for NH are currently limited to temperatures of a few
hundred mK. Producing NH molecules in the ultracold regime requires a
second-stage cooling mechanism, e.g.\ sympathetic cooling with an ultracold
atomic gas \cite{soldan:09,wallis:09,wallis:10,zuchowski:11,hummon:11} or
evaporative cooling. The latter process relies on elastic, thermalizing NH + NH
collisions as the magnetic trap depth is gradually reduced. Inelastic
spin-changing collisions lead to immediate trap loss and are therefore
unfavorable. It is generally accepted that, in order to achieve 
evaporative cooling, elastic collisions should be much more efficient than
inelastic ones. More specifically, a Monte Carlo study on evaporative cooling of cesium atoms
indicated that the ratio between elastic and inelastic collision rates should be greater than
150 \cite{monroe:93}. Although evaporative cooling of NH might 
work with a lower ratio, it will be assumed that 150 is also the minimum
required value for NH + NH collisions.


For two magnetically trapped NH($X\,^3\Sigma^-$) molecules, the collision
complex is in the low-field seeking $| S=2, M_S=2 \rangle$ quintet spin state,
with $S$ denoting the total electronic spin and $M_S$ its projection on the
magnetic-field axis. Inelastic transitions may change either the $M_S$ quantum
number of the quintet state, or the total spin $S$ to produce singlet or triplet complexes. The $S=0$
and 1 dimer spin states are chemically reactive
\cite{dhont:05,janssen:09,lai:03} and 
could be of interest in cold controlled chemistry experiments \cite{krems:08}.

A rigorous calculation of elastic and inelastic cross sections requires a full
quantum coupled-channels method. In the case of NH--NH, however, the strong 
anisotropy of the interaction potentials and the open-shell nature of the monomers
gives rise to a very large number of channels, making the calculation extremely
challenging.
In a recent study by Tscherbul \textit{et al.}\ \cite{tscherbul:09c} on the iso-electronic
O$_2$($X\,^3\Sigma^-_g$) -- O$_2$($X\,^3\Sigma^-_g$) system, close-coupling
calculations were performed that included up to 2526 channels, yielding cross
sections converged to within 10\%. These calculations
were carried out in a fully decoupled channel basis to study
collisions in the presence of an external magnetic field.
It was noted, however,
that the true O$_2$--O$_2$ interaction potential is likely to be more
anisotropic than the potential used in their work, thus implying that even more
channels would be needed. Other quantum scattering studies on O$_2$--O$_2$ include 
those by Avdeenkov and Bohn \cite{avdeenkov:01} and P\'{e}rez-R\'{i}os
\textit{et al.}\ \cite{perez-rios:09}.
In the work of Avdeenkov and Bohn, field-free collisions were studied using a
total angular momentum representation, thereby reducing the total number of
channels to 836. The rotational basis-set size used in these calculations
was, however, smaller than that used in Ref.\ \cite{tscherbul:09c}.
P\'{e}rez-R\'{i}os \textit{et al.}\ also employed a total angular momentum
basis, but the O$_2$ monomers were treated as closed-shell molecules. This allowed 
them to reduce the number of channels to 300.%

To our knowledge, only one theoretical study has been reported for the NH--NH
system.  Kajita \cite{kajita:06} employed the Born approximation,
distorted-wave Born approximation, and classical path method to calculate elastic
and inelastic cross sections at energies ranging from 1 $\mu$K to 10 K, and
found that evaporative cooling of NH is likely to be feasible.  It must be
noted, however, that only the electric dipole-dipole and magnetic
dipole-dipole interactions were considered in these calculations.

The aim of the present work is to obtain cold and ultracold NH + NH collision
cross sections from rigorous quantum scattering calculations on an accurate
\textit{ab initio} quintet potential-energy surface.
We include intramolecular spin-spin, spin-rotation, and intermolecular magnetic
dipole-dipole coupling in the dynamics.
In addition, we seek to address the issue of
dealing with very large basis sets in order to converge the scattering results, a problem
that is general for open-shell systems with relatively deep potential energy wells. 
For this purpose, we have employed a total angular momentum representation to perform the scattering
calculations, assuming zero field.
Collisions in a magnetic field are discussed in a separate publication \cite{janssen:11a}.
It will be shown that, within the uncertainty
limits of the interaction potential, even an unconverged basis set can provide
meaningful results. 

This paper is organized as follows. In Sec.\ \ref{sec:theory}, we discuss
the scattering Hamiltonian and channel basis-set functions, followed by the details of the
cross-section calculations. Results are presented in Sec.\ \ref{sec:results_A}.
In Secs.\ \ref{sec:results_B} and \ref{sec:results_C}, we provide a
comprehensive discussion on the accuracy of our calculated cross sections.
Conclusive remarks are given in Sec.\ \ref{sec:concl}.

\section{Theory}
\label{sec:theory}
\subsection{Hamiltonian and channel basis functions}
\label{subsec:H}
We consider the case of two colliding NH($^3\Sigma^-$) molecules in the absence
of an external field and treat the monomers as rigid rotors.
We use a space-fixed coordinate frame to describe the collision complex. The
relevant Jacobi coordinates are the intermolecular vector $\bm{R}$ that
connects the centers of mass of molecules $A$ and $B$, and the polar angles
$\omega_i = (\theta_i,\phi_i)$ of the two monomers ($i=A,B$).  We will neglect
hyperfine coupling and assume that both monomers are in their nuclear-spin
stretched states $| I, M_I=I \rangle$, with $I=I_{\rm{N}}+I_{\rm{H}}$ denoting the maximum total
nuclear spin and $M_I$ its laboratory-frame projection. For fermionic $^{14}$NH
the maximum nuclear spin is $I=3/2$ and for bosonic $^{15}$NH we have $I=1$.

The scattering Hamiltonian for NH--NH can be written as 
\begin{equation}
\label{eq:H}
\hat{H} = -\frac{\hbar^2}{2\mu R} \frac{\partial^2}{\partial R^2}R +
          \frac{\hat{L}^2}{2\mu R^2} +
          V_S(\bm{R},\omega_A,\omega_B) +
          V_{\rm{magn.dip}}(\bm{R},\hat{\bm{S}}_A,\hat{\bm{S}}_B) +
          \hat{H}_A + \hat{H}_B,
\end{equation}
where $\mu$ is the reduced mass of the complex, $R$ is the length of the
vector $\bm{R}$, $\hat{L}^2$ is the angular momentum operator associated with
rotation of $\bm{R}$, $V_S(\bm{R},\omega_A,\omega_B)$ is the potential-energy
surface for total spin $S$, 
$V_{\rm{magn.dip}}(\bm{R},\hat{\bm{S}}_A,\hat{\bm{S}}_B)$ is the intermolecular magnetic dipole
interaction between the two triplet spins, and $\hat{H}_A$ and $\hat{H}_B$ are
the Hamiltonians of the individual monomers. The magnetic dipole term is given
by
\begin{equation}
V_{\rm{magn.dip}}(\bm{R},\hat{\bm{S}}_A,\hat{\bm{S}}_B) = - \sqrt{6} g_S^2 \mu_{\rm{B}}^2 \frac{\alpha^2}{R^3}
                  \sum_q (-1)^q C_{2,-q}(\Omega) [\hat{\bm{S}}_A \otimes \hat{\bm{S}}_B]_q^{(2)},
\end{equation}
where $g_S \approx 2.0023$ is the electron $g$-factor,
$\mu_{\rm{B}}$ is the Bohr magneton, $\alpha$ is the fine-structure constant, $C_{2,-q}$ is a
Racah-normalized spherical harmonic,
$\Omega = (\Theta,\Phi)$ describes the orientation of $\bm{R}$ in the space-fixed frame,
and the factor in square brackets represents 
the tensorial product of the monomer spin operators $\hat{\bm{S}}_A$ and $\hat{\bm{S}}_B$.
The monomer operators $\hat{H}_i$ each contain a rotation, spin-rotation, and
intramolecular spin-spin term:
\begin{equation}
\label{eq:Hi}
\hat{H}_i = B_0 \hat{N}_i^2 + \gamma\hat{\bm{N}}_i\cdot\hat{\bm{S}}_i +
            \frac{2}{3}\sqrt{6} \lambda_{\rm{SS}}
            \sum_q (-1)^q C_{2,-q}(\omega_i) [\hat{\bm{S}}_i \otimes \hat{\bm{S}}_i]_q^{(2)},
\end{equation}
with $\hat{N}_i$ denoting the rotational angular momentum operator of monomer
$i$. For brevity, we will denote the intramolecular spin-spin operator as
$\hat{V}_{\rm{SS}}^{(i)}$.  The numerical values for the rotational, spin-rotation,
and spin-spin constants are $B_0 = 16.343275$ cm$^{-1}$, $\gamma = -0.05486$
cm$^{-1}$, and $\lambda_{\rm{SS}} = 0.91989$ cm$^{-1}$ \cite{ram:10} for $^{14}$NH,
and, by scaling with the isotope mass (see e.g.\ p.\ 239 of Ref.\
\cite{mizushima:75}), we obtain $B_0 = 16.270340$ cm$^{-1}$, $\gamma = -0.05460$
cm$^{-1}$, and $\lambda_{\rm{SS}} = 0.91989$ cm$^{-1}$ for $^{15}$NH.

For the interaction potential $V_S(\bm{R},\omega_A,\omega_B)$ we take the $S=2$
\textit{ab initio} surface of Ref.\ \cite{janssen:09}. This spin state
corresponds to the case where both molecules are in their magnetically trapped
(spin-stretched) states.  Although the potential is based on the Jacobi
coordinates for $^{14}$NH -- $^{14}$NH, we use the same surface for the
$^{15}$NH -- $^{15}$NH isotope. This approximation is very reasonable since the
center of mass of $^{15}$NH is shifted by only 0.008 $a_0$ with respect to that
of $^{14}$NH.  We have verified that, at the equilibrium distance of the
complex, this would give a maximum error of 2.2\% in the $^{15}$NH -- $^{15}$NH
potential, which falls within the uncertainty range of the \textit{ab initio}
data.  Following Ref.\ \cite{green:75}, we expand the quintet potential in
terms of spherical harmonics $Y_{L,M}$ of degree $L$ and order $M$:
\begin{align}
\label{eq:V_sf}
V(\bm{R},\omega_A,\omega_B) = {} & \sum_{L_A,L_B,L_{AB}} \upsilon_{L_A,L_B,L_{AB}}(R)
                              A_{L_A,L_B,L_{AB}}(\Omega,\omega_A,\omega_B), \\
\label{eq:A_LLL}
A_{L_A,L_B,L_{AB}}(\Omega,\omega_A,\omega_B) = {} & \sum_{M_A,M_B,M_{AB}}
                            \langle L_A M_A L_B M_B | L_{AB} M_{AB} \rangle \nonumber \\
 & \times  Y_{L_A,M_A}(\omega_A) Y_{L_B,M_B}(\omega_B)  Y^*_{L_{AB},M_{AB}}(\Omega), 
\end{align}
where $\langle L_A M_A L_B M_B | L_{AB} M_{AB} \rangle$ is a Clebsch-Gordan coefficient 
and the superscript * denotes complex conjugation.
The subscript $S=2$ has been omitted for brevity. 
It should be noted that
the angular functions of Eq.\ (\ref{eq:A_LLL}) differ by a
factor of $\zeta = (-1)^{L_A-L_B} (4\pi)^{-3/2} (2L_{AB}+1) [(2L_A+1)(2L_B+1)]^{1/2}$
from the functions used in Ref.\ \cite{janssen:09}, i.e.\ the $\upsilon_{L_A,L_B,L_{AB}}(R)$
expansion coefficients of Ref.\ \cite{janssen:09} must be multiplied
by $\zeta$ to obtain the potential in the form of Eq.\ (\ref{eq:V_sf}).

In the absence of an external field, both the total angular momentum $\mathcal{J}$
and its space-fixed projection $\mathcal{M}$ are rigorously conserved.
We therefore expand the wave function in a total angular momentum basis:
\begin{align}
\label{eq:basis_tot}
\Psi^{\mathcal{J},\mathcal{M}}(R,\Omega,\omega_A,\omega_B,\sigma_A,\sigma_B) = {} &
\frac{1}{R} \sum_{N_A,N_B,N,S_A,S_B,S,J,L}
\chi^{\mathcal{J},\mathcal{M}}_{N_A,N_B,N,S_A,S_B,S,J,L}(R)
\nonumber \\
 & \times \psi^{\mathcal{J},\mathcal{M}}_{N_A,N_B,N,S_A,S_B,S,J,L}(\Omega,\omega_A,\omega_B,\sigma_A,\sigma_B),
\end{align}
where $\sigma_A$ and $\sigma_B$ refer to the electronic spin coordinates of molecules $A$ and $B$, respectively.
Here $N_A$ and $N_B$ denote the rotational quantum numbers of the two monomers, $N$ is the 
coupled rotational quantum number of the complex, $S_A$ and $S_B$ are the monomer spin quantum numbers, 
which are coupled into total spin $S$,
$J$ is the
angular momentum quantum number arising from the coupling of $N$ and $S$, and $L$ denotes the partial-wave
angular momentum.
The coupled angular momentum basis functions are defined as follows:
\begin{eqnarray}
\label{eq:basis_ang}
\lefteqn{\psi^{\mathcal{J},\mathcal{M}}_{N_A,N_B,N,S_A,S_B,S,J,L}(\Omega,\omega_A,\omega_B,\sigma_A,\sigma_B) =} \nonumber \\
& & \sum_{M_J,M_L} \sum_{M_N,M_S}  \sum_{M_{S_A},M_{S_B}} \sum_{M_{N_A},M_{N_B}}
Y_{N_A,M_{N_A}}(\omega_A) Y_{N_B,M_{N_B}}(\omega_B) Y_{L,M_L}(\Omega)        \nonumber \\
& & \times \tau_{S_A,M_{S_A}}(\sigma_A) \tau_{S_B,M_{S_B}}(\sigma_B) 
\langle N_A M_{N_A} N_B M_{N_B} | N M_N \rangle
\langle S_A M_{S_A} S_B M_{S_B} | S M_S \rangle   \nonumber \\
& & \times \langle N M_N S M_S | J M_J \rangle
\langle J M_J L M_L | \mathcal{J} \mathcal{M} \rangle,
\end{eqnarray}
where $\tau_{S_A,M_{S_A}}$ and $\tau_{S_B,M_{S_B}}$ are spinor wave functions. Here the quantum numbers
$M_{N_i}$, $M_{S_i}$, $M_{N}$, $M_S$, $M_{J}$, and $M_{L}$ denote the projections of
$N_i$, $S_i$, $N$, $S$, $J$, and $L$ onto the magnetic-field axis.
We will restrict the basis such that $N_A$ and $N_B$ range from 0 to $N_{\rm{max}}$ and
$L = 0, \hdots, L_{\rm{max}}$.
Note that the scattering calculations in this basis may also be performed for a
single dimer spin state $S$. 
As detailed in Section \ref{sec:results_B}, we will exploit this feature to investigate
the validity of describing all three dimer spin states by the $S=2$ potential energy surface.

Since target and
projectile are identical, we can symmetrize the wave function with respect to the permutation operator
$\hat{P}_{AB}$. This yields the following normalized basis functions:
\begin{eqnarray}
\label{eq:basis_sa}
\lefteqn{\phi^{\eta,\mathcal{J},\mathcal{M}}_{N_A,N_B,N,S_A,S_B,S,J,L} = 
\frac{1}{[2(1+\delta_{N_A N_B}\delta_{S_A S_B})]^{1/2}}  
\big[ \psi^{\mathcal{J},\mathcal{M}}_{N_A,N_B,N,S_A,S_B,S,J,L} } \nonumber \\
 & & \hphantom{X} + \eta (-1)^{L+N_A+N_B-N+S_A+S_B-S}
\psi^{\mathcal{J},\mathcal{M}}_{N_B,N_A,N,S_B,S_A,S,J,L} \big]. 
\hphantom{XXXXXX}
\end{eqnarray}
Here $\eta=+1$ corresponds to composite bosons and $\eta=-1$ to composite
fermions, assuming that the molecules are in their nuclear-spin stretched states.
To obtain a linearly independent basis, the index pair ($N_A,N_B$)
must be restricted such that $N_A \geq N_B$ 
\cite{green:75}. Finally, the basis functions of Eq.\ (\ref{eq:basis_sa}) are
also eigenfunctions of the inversion operator, with eigenvalues
$\epsilon = (-1)^{N_A+N_B+L}$.  Thus, the Hamiltonian in the symmetry-adapted basis
consists of four blocks, each block labeled by $\eta$ and the parity
$\epsilon$. It must be noted, however, that the wave function of Eq.\ (\ref{eq:basis_sa})
vanishes for ($\eta=+1, \epsilon=-1$) and ($\eta=-1, \epsilon=+1$) if the
molecules are in the magnetically trapped 
ground state with $N_A=N_B=N=0$ and $S=2$.
We therefore only need to consider the parity case $\epsilon=+1$ for $\eta=+1$ and
$\epsilon=-1$ for $\eta=-1$.

The matrix elements of the Hamiltonian in the symmetry-adapted basis [Eq.\
(\ref{eq:basis_sa})] can be readily obtained from the matrix elements in the
`primitive' basis $\psi^{\mathcal{J},\mathcal{M}}_{N_A,N_B,N,S_A,S_B,S,J,L}$.
These are given in the Appendix.

\subsection{$S$-matrices and cross sections}
The close-coupling equations are solved for each $\mathcal{J}$ and each
symmetry type ($\eta, \epsilon$) using the hybrid log-derivative method of
Alexander and Manolopoulos \cite{alexander:87}. This algorithm uses a fixed-step-size
log-derivative propagator in the short range and a variable-step-size Airy
propagator in the long range. The solutions are then matched to asymptotic
boundary conditions to obtain the scattering $S$-matrices. Since we consider
only the field-free case, the results are independent of the total
angular momentum projection $\mathcal{M}$.

Although we assume zero magnetic field in our calculations, we are ultimately
interested in the elastic and inelastic spin-changing cross sections for
magnetically trapped NH. It is therefore necessary to transform the
$S$-matrices to a channel product eigenbasis of the form $|
(\bar{N}_A,S_A)J_A, M_{J_A}\rangle | (\bar{N}_B,S_B)J_B, M_{J_B}\rangle |
L, M_L \rangle$, where $J_i$ and $M_{J_i}$ arise from the angular momentum
coupling of $\bar{N}_i$ and $S_i$. Here we have used the notation
$\bar{N}_i$ instead of $N_i$, because $N_i$ is strictly not a good quantum number.
This is due to the intramolecular spin-spin coupling, which mixes states with $N_i$ and $N_i\pm 2$.
However, the mixing is quite weak and $\bar{N_i}$ corresponds almost exactly to $N_i$.
A symmetry-adapted version of the channel eigenbasis is given by
\begin{eqnarray}
\label{eq:eigenbasis_sa}
\lefteqn{\phi^{\eta}_{\bar{N}_A,S_A,J_A,M_{J_A},\bar{N}_B,S_B,J_B,M_{J_B},L,M_L} = 
\frac{1}{[2(1+\delta_{\bar{N}_A \bar{N}_B}\delta_{S_A S_B}\delta_{J_A J_B}\delta_{M_{J_A} M_{J_B}} )]^{1/2}} } \nonumber \\
& & \hphantom{X} \times \big[ | (\bar{N}_A,S_A)J_A, M_{J_A}\rangle | (\bar{N}_B,S_B)J_B, M_{J_B}\rangle | L, M_L \rangle  \nonumber \\
& & \hphantom{X} \times + \eta (-1)^L
    | (\bar{N}_B,S_B)J_B, M_{J_B}\rangle | (\bar{N}_A,S_A)J_A, M_{J_A}\rangle | L, M_L \rangle \big].
\hphantom{XXXX}
\end{eqnarray}
It should be noted that the total angular momentum $\mathcal{J}$ is not a good
quantum number here, but its laboratory-frame projection $\mathcal{M} = M_{J_A}
+ M_{J_B} + M_L$ is conserved. 

The basis transformation from Eq.\
(\ref{eq:basis_sa}) to Eq.\ (\ref{eq:eigenbasis_sa}) cannot be performed
analytically, because $N_i$, $N$, and $S$ are only approximately good quantum
numbers.  We have therefore developed a numerical scheme in which the channel
eigenfunctions of Eq.\ (\ref{eq:eigenbasis_sa}) are obtained as the
simultaneous eigenvectors of the operators $\{\hat{L}^2, \hat{H}_A+\hat{H}_B,
\hat{J}_{z_A}+\hat{J}_{z_B},\hat{J}^2_{z_A}+\hat{J}^2_{z_B}\}$.  Note that
these operators all commute with each other and with $\hat{P}_{AB}$.  The
numerical procedure works as follows. We start by diagonalizing the first
operator, e.g.\ the matrix representation of the $\hat{L}^2$ operator,
constructed in the basis of Eq.\ (\ref{eq:basis_sa}).  In each degenerate
subspace of $\hat{L}^2$, we set up the matrix of the next operator and
diagonalize it. This process is repeated for the remaining operators until all
eigenvectors are unique. We note that the operator
$\hat{J}^2_{z_A}+\hat{J}^2_{z_B}$ is only required to distinguish between
states with coincidental degeneracies in $M_{J_A} + M_{J_B}$, e.g.\ the states
$| \phi^{\eta}_{0,1,1,0,0,1,1,0,0,0} \rangle$ with $M_{J_A}=M_{J_B}=0$ and $|
\phi^{\eta}_{0,1,1,1,0,1,1,-1,0,0} \rangle$ with $M_{J_A}=1, M_{J_B}=-1$. 
Any
remaining degeneracies arising from $\hat{H}_A+\hat{H}_B$ may be lifted by
diagonalizing the operator $\hat{H}_A^2 + \hat{H}_B^2$, but such degeneracies
occur only for higher energies. In the cold and ultracold regime, these higher-energy channels
are closed and the
eigenvalues of $\hat{L}^2$, $\hat{H}_A+\hat{H}_B$,
$\hat{J}_{z_A}+\hat{J}_{z_B}$, and $\hat{J}^2_{z_A}+\hat{J}^2_{z_B}$ are
sufficient to uniquely identify all relevant quantum numbers. It must be noted
that, since $\hat{J}_{z_A}$ and $\hat{J}_{z_B}$ do not separately commute with
$\hat{P}_{AB}$, the matrices of $\hat{J}_{z_A}+\hat{J}_{z_B}$ and
$\hat{J}^2_{z_A}+\hat{J}^2_{z_B}$ are not trivially constructed in the basis of
Eq.\ (\ref{eq:basis_sa}).  We obtained these matrices by first evaluating the
$\hat{J}_{z_i}$ and $\hat{J}^2_{z_i}$ operators in a fully decoupled basis of
the form $| N_A, M_{N_A}, S_A, M_{S_A}, N_B, M_{N_B}, S_B, M_{S_B}, L, M_L
\rangle$.  Both $\hat{J}_{z_i}$ and $\hat{J}^2_{z_i}$ are diagonal in this
basis, with diagonal elements $M_{J_i} = M_{N_i} + M_{S_i}$ and $M_{J_i}^2$,
respectively.  We subsequently performed an analytical transformation to the
coupled basis of Eq.\ (\ref{eq:basis_ang}) using the appropriate Clebsch-Gordan
coefficients. Finally, we used a rectangular transformation matrix for
$\hat{J}_{z_A}+\hat{J}_{z_B}$ and $\hat{J}^2_{z_A}+\hat{J}^2_{z_B}$ to account
for the symmetry adaptation, i.e.\ to transform the matrices to the basis of
Eq.\ (\ref{eq:basis_sa}).

The evaporative cooling rate for cold magnetically trapped NH molecules, with
quantum numbers $\bar{N}_A=\bar{N}_B=0$, $J_A=J_B=1$, and $M_{J_A}=M_{J_B}=1$, is
determined by the ratio between elastic and $M_J$-changing cross sections. The
cross-section expression for indistinguishable molecules at total energy $E$ is
\cite{tscherbul:09c}
\begin{equation}
\label{eq:xsec}
\sigma^{\eta}_{\gamma_A \gamma_B \rightarrow \gamma'_A \gamma'_B}(E) =
\frac{\pi (1+\delta_{\gamma_A \gamma_B})}{k^2_{\gamma_A \gamma_B}}
\sum_{L,M_L} \sum_{L', M'_L}
\left| T^{\eta}_{\gamma_A \gamma_B L M_L;\gamma'_A \gamma'_B L' M'_L}(E) \right|^2,
\end{equation}
where we have introduced the shorthand notation $\gamma_A\gamma_B$ to label the 
symmetrized monomer states, i.e.\ $\phi^{\eta}_{\bar{N}_A,S_A,J_A,M_{J_A},\bar{N}_B,S_B,J_B,M_{J_B},L,M_L}
\equiv | \gamma_A \gamma_B \rangle | L M_L \rangle$, and $k_{\gamma_A \gamma_B}$ is the
length of the wavevector for the initial collision channel $| \gamma_A \gamma_B \rangle$.
The $T$-matrix elements are defined in terms of the transformed $S$-matrix elements
as $T^{\eta}_{\gamma_A \gamma_B L M_L;\gamma'_A \gamma'_B L' M'_L}
= \delta_{\gamma_A \gamma'_A} \delta_{\gamma_B \gamma'_B} \delta_{L L'} \delta_{M_L M_L'} -
S^{\eta}_{\gamma_A \gamma_B L M_L;\gamma'_A \gamma'_B L' M'_L}$.
Finally, we note that the summations over $M_L$ and $M'_L$ in Eq.\ (\ref{eq:xsec})
may also be understood as a sum over all possible $\mathcal{M}$ values, since
$\mathcal{M} = M_{J_A}+ M_{J_B} + M_L = M'_{J_A}+ M'_{J_B} + M'_L$.

\subsection{Computational details}
The scattering calculations were performed using a modified version of the
MOLSCAT package \cite{molscat:94,gonzalez:07} in which the coupled basis set of Eq.\
(\ref{eq:basis_tot}) was implemented.  The radial grid ranged from $4.5$ to
$500$ $a_0$, with the Airy propagation starting at $15$ $a_0$. The step
size for the log-derivative propagator was 0.02 $a_0$.  The basis set included
all functions up to $N_A = N_B = 5$ and $L = 6$.
The expansion of the quintet potential was truncated at $L_A = L_B = 6$.
As mentioned in Section
\ref{subsec:H}, the chemically reactive singlet and triplet interaction
potentials were excluded from the calculations, and were replaced by the
nonreactive $S=2$ surface. Thus, we assumed that all three spin states are
described by the same potential energy surface.
In order to study the role of the $S=0$ and 1 states under this assumption,
we also performed scattering calculations for the quintet state only.

At each collision energy, the scattering $S$-matrices were accumulated for all relevant
$\mathcal{J}$ values and subsequently transformed to the channel eigenbasis
of Eq.\ (\ref{eq:eigenbasis_sa}) for all possible $\mathcal{M}$ values.
The basis transformation was carried out in Matlab \cite{matlab:09}. The total elastic
and inelastic cross sections were then obtained using Eq.\ (\ref{eq:xsec}).

\section{Results and discussion}
\label{sec:results}

\subsection{Cross sections}
\label{sec:results_A}

\begin{figure}
\centering
\includegraphics[width=15cm]{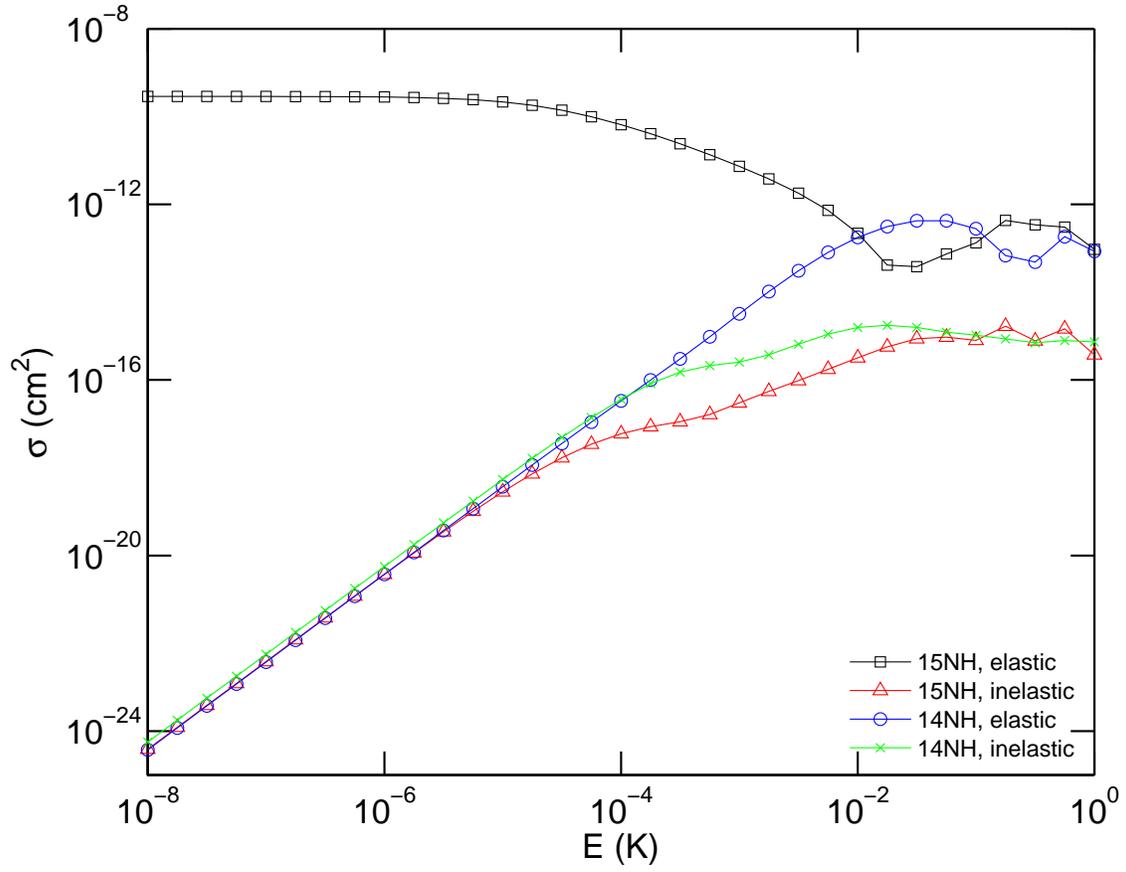} 
\caption
  {\label{fig:xsec1415}
Elastic and inelastic $M_{J}$-changing cross sections for $^{14}$NH + $^{14}$NH and
$^{15}$NH + $^{15}$NH collisions, assuming that all molecules are in their magnetically
trappable and nuclear-spin stretched state.
}
\end{figure}

The elastic and $M_J$-changing cross sections for magnetically trapped
$^{14}$NH and $^{15}$NH are shown in Fig.\ \ref{fig:xsec1415}.  At low collision
energies, the cross sections are dominated by incoming $s$-waves for bosonic $^{15}$NH
and by $p$-waves for fermionic $^{14}$NH. The observed energy dependence is consistent
with Wigner's threshold law for iso-energetic processes \cite{wigner:48,krems:03b}:
\begin{equation}
\label{eq:Wigner}
\sigma \propto E^{L+L'},
\end{equation}
where $L$ and $L'$ denote the partial waves in the incoming and outgoing channels, respectively.
For elastic $^{15}$NH + $^{15}$NH collisions, we have $L=L'=0$ and the cross section is
constant as a function of $E$. For inelastic collisions, the change in $M_{J_A}$ or $M_{J_B}$ 
must be accompanied by a change in the $M_L$ quantum number, which follows from the conservation of $\mathcal{M}$.
Since the parity $(-1)^{N_A+N_B+L}$ is also rigorously conserved, it is easily verified
[see Eq.\ (\ref{eq:basis_sa})]
that the dominant inelastic cross section for $^{15}$NH ($L=0$) corresponds to the $L'=2$ outgoing
channel, and consequently behaves as $E^2$.
For fermionic
$^{14}$NH + $^{14}$NH collisions, both the elastic and inelastic channels are dominated
by $L=L'=1$ [see Eq.\ (\ref{eq:basis_sa})], yielding the observed $E^2$ behaviour.
We also point out that, in the presence of a magnetic field, all inelastic transitions
would be exothermic and the corresponding cross section would behave as $E^{L-1/2}$ \cite{wigner:48}.
This leads to a different elastic-to-inelastic collision ratio than in the field-free case.
It is shown in a separate publication that
the ratio for $^{15}$NH + $^{15}$NH collisions is still very favorable 
when the magnetic field is explicitly included \cite{janssen:11a}.

\begin{figure}
\centering
\includegraphics[width=15cm]{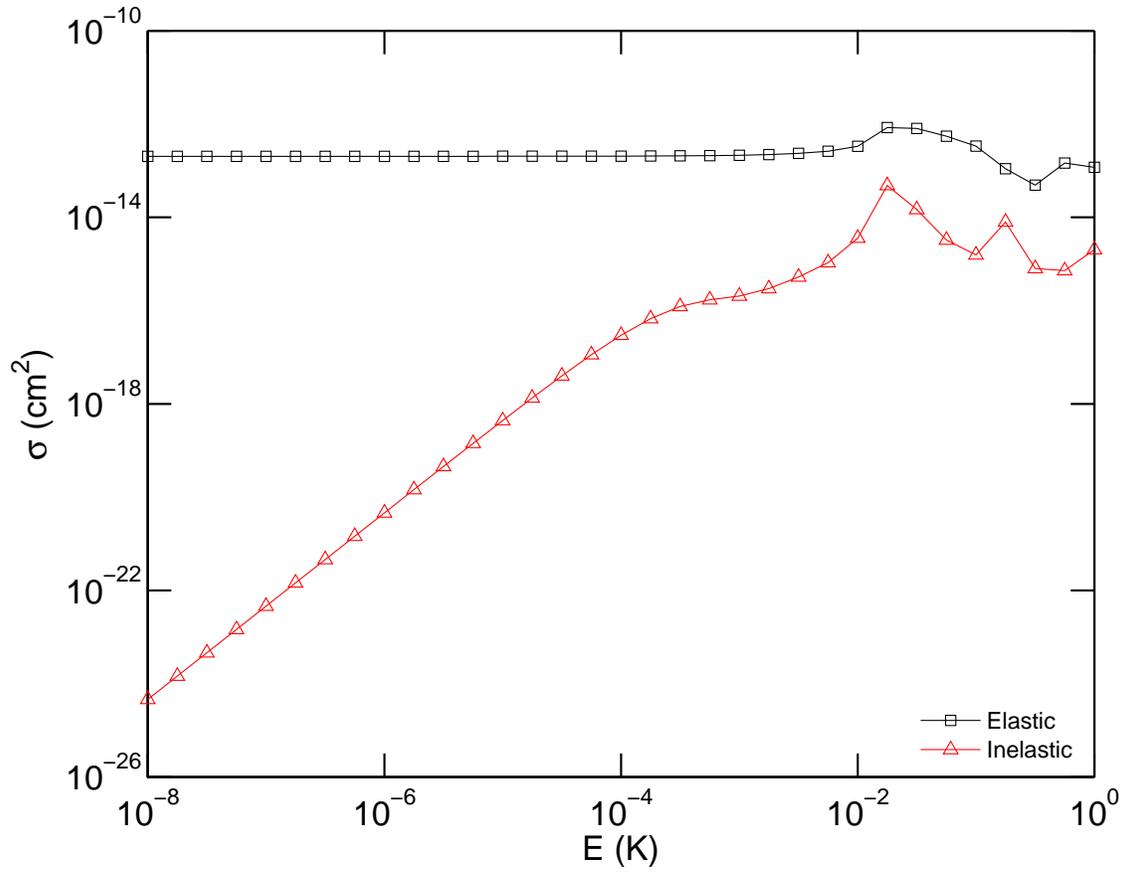} 
\caption
  {\label{fig:14NH_spinstat}
Elastic and inelastic $M_{J}$-changing cross sections for magnetically trapped $^{14}$NH,
assuming a statistical mixture of nuclear-spin states.
}
\end{figure}

We find that $^{15}$NH is more suitable for evaporative cooling than
$^{14}$NH, in agreement with the findings of Kajita \cite{kajita:06}.
More specifically, we see in Fig.\ \ref{fig:xsec1415}
 that the elastic-to-inelastic ratio for $^{15}$NH + $^{15}$NH
far exceeds the critical value of 150 for all energies below $E\approx 10^{-2}$ K, 
while for $^{14}$NH + $^{14}$NH the ratio is orders of magnitude smaller
and is close to unity at collision energies below $10^{-4}$ K.
This result is essentially a consequence of the Pauli principle,
which forbids $s$-wave scattering for $^{14}$NH + $^{14}$NH.
We emphasize that our calculations were performed under the assumption that both
molecules are in their nuclear-spin stretched states, giving rise to a
symmetric nuclear-spin wave function. This leads to the restriction that
$\eta=+1$ ($\epsilon=+1$) for $^{15}$NH and $\eta=-1$ ($\epsilon=-1$) for
$^{14}$NH.  If, however, the two monomers were in \textit{different}
nuclear-spin states, the corresponding wave function may also be antisymmetric
under exchange and both values of $\eta$ would be allowed.  In that case, the
total cross section is given by a weighted sum over the cross sections
$\sigma^{+1}$ and $\sigma^{-1}$:
\begin{equation}
\sigma_{\gamma_A \gamma_B \rightarrow \gamma'_A \gamma'_B}(E) = 
W^+ \sigma^{+1}_{\gamma_A \gamma_B \rightarrow \gamma'_A \gamma'_B}(E) + 
W^- \sigma^{-1}_{\gamma_A \gamma_B \rightarrow \gamma'_A \gamma'_B}(E),
\end{equation}
with $W^+$ and $W^-$ denoting the relative spin-statistical weights. 
The weights are $W^+=5/12$ and $W^-=7/12$ for fermionic $^{14}$NH and 
3/4 and 1/4 for bosonic $^{15}$NH.
Figure \ref{fig:14NH_spinstat} shows the results for $^{14}$NH
-- $^{14}$NH, assuming a mixture of different nuclear-spin states.  The
inclusion of even-$L$ partial waves ($\eta=+1$) strongly enhances the efficiency
of evaporative cooling for $^{14}$NH, in particular due to the $s$-wave elastic
contribution.
For $^{15}$NH -- $^{15}$NH, the addition of odd-$L$ partial wave contributions ($\eta=-1$)
will probably lead to a slightly lower elastic-to-inelastic ratio. This is because
the odd-$L$ elastic cross section, which vanishes as $E^2$, is almost negligible compared to the 
$s$-wave elastic cross section in the ultracold limit. The odd-$L$ inelastic contribution,
on the other hand, exhibits the same threshold behaviour as the even-$L$ inelastic cross section,
and could easily increase the total inelastic loss by a factor of $\sim 2$.
Hence we conclude
that, in order to achieve efficient evaporative cooling, bosonic $^{15}$NH
should be prepared in a single nuclear-spin state, 
while for $^{14}$NH the molecules should be in a mixture of hyperfine states. 


Aside from symmetry arguments, the difference between $^{15}$NH -- $^{15}$NH
and $^{14}$NH -- $^{14}$NH is relatively small. The rotational and
spin-rotation constants differ by only 0.45\% and the reduced masses of the
collision complex are 6.6\% different. Since $^{15}$NH is more advantageous for
evaporative cooling, we will only consider collisions between $^{15}$NH
molecules in the remainder of this work. Again it will be assumed that the
monomers are in identical hyperfine states, so that only the $\eta=+1$
($\epsilon=+1$) symmetry case needs to be examined.

\begin{figure}
\centering
\includegraphics[width=15cm]{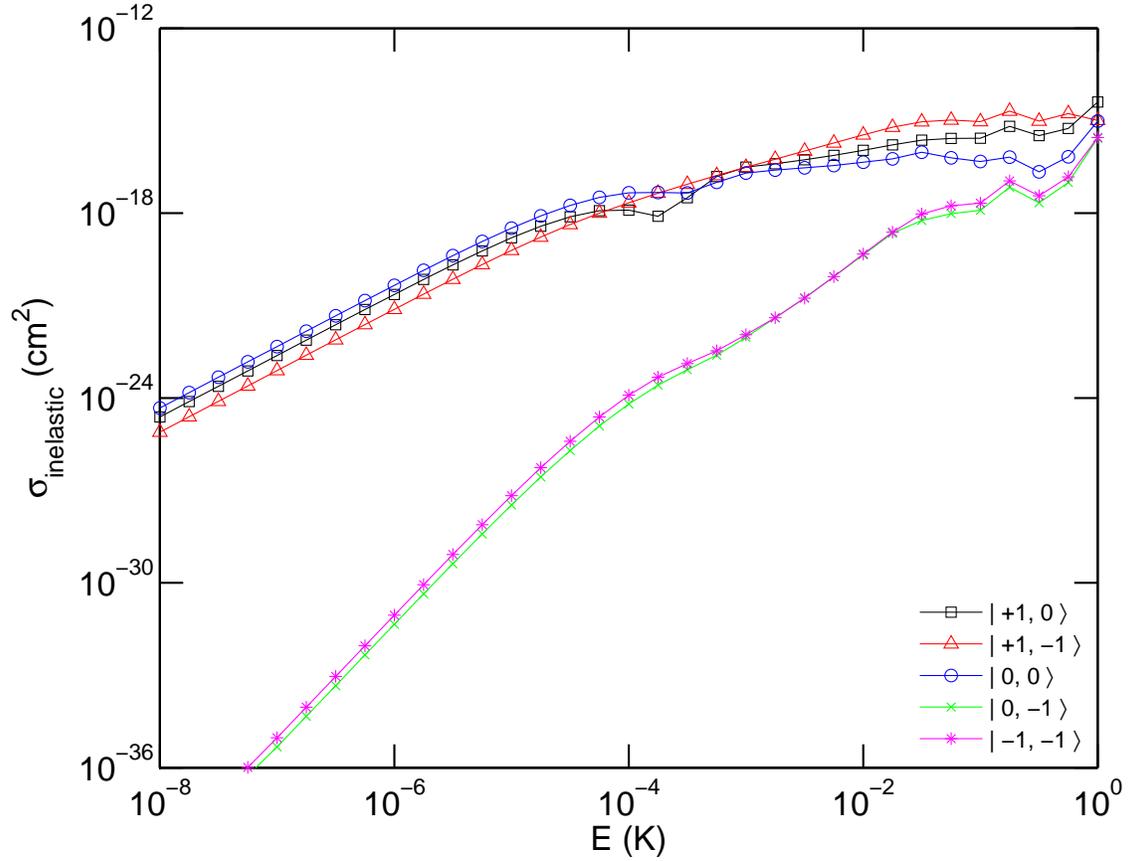} 
\caption
  {\label{fig:xsec_ss}
State-to-state inelastic cross sections for magnetically trapped $^{15}$NH as a function
of collision energy. The final states are labeled by $| M_{J_A}, M_{J_B} \rangle$.
}
\end{figure}

State-to-state inelastic cross sections for magnetically trapped $^{15}$NH ($M_{J_A}=1,M_{J_B}=1$)
are shown in Fig.\ \ref{fig:xsec_ss}. We find that transitions 
to the states with $|M_{J_A}=1,M_{J_B}=0 \rangle$,
$|M_{J_A}=1,M_{J_B}=-1 \rangle$, and $|M_{J_A}=0,M_{J_B}=0 \rangle$ are
dominant in the ultracold regime. It can also be seen that these cross sections
follow an $E^2$ dependence below $\sim10^{-4}$ K.  The inelastic cross sections for
$|M_{J_A}=0,M_{J_B}=-1 \rangle$ and $|M_{J_A}=-1,M_{J_B}=-1 \rangle$ exhibit
$E^4$ behaviour at low collision energies. These results are consistent with
the threshold laws of Krems and Dalgarno for collisional reorientation of
angular momentum in the absence of an external field \cite{krems:03b}. Although these
laws were derived for collisions of paramagnetic species with structureless targets, they
also apply to $^{15}$NH + $^{15}$NH collisions:
\begin{equation}
\label{eq:deltaMeven}
\sigma_{J,M_J \rightarrow J,M_J\pm\Delta M_J} \propto E^{\Delta M_J}
\end{equation}
if $\Delta M_J$ is even and
\begin{equation}
\label{eq:deltaModd}
\sigma_{J,M_J \rightarrow J,M_J\pm\Delta M_J} \propto E^{\Delta M_J+1}
\end{equation}
if $\Delta M_J$ is odd. Here $\Delta M_J$ is defined as the change in $M_{J_A} + M_{J_B}$.
It also follows from Eq.\ (\ref{eq:deltaMeven}) that the elastic cross section
($\Delta M_J = 0$) is constant at low energies, in agreement with Eq.\ (\ref{eq:Wigner}).


\subsection{Contributions from singlet and triplet states}
\label{sec:results_B}

Throughout this paper, we have assumed that all three spin states of the NH--NH
complex are described by a single nonreactive potential-energy surface, namely
the $S=2$ surface.  The $S=2$ state corresponds to the case where both monomers
are magnetically trapped, and is therefore the most relevant spin state in our
present study. It is, however, not \textit{a priori} clear how the $S=0$ and 1
states can influence the trap loss probability, and how well they can be described
by the quintet surface.

We must first point out that, even at infinite separation,
$S$ is strictly not a good quantum number due to the intramolecular spin-spin coupling.
However, the coupling between different spin states is relatively weak and we may
therefore treat $S$ as nearly exact. 
Specifically, for the rotational ground state of the complex, the initial state with $M_{J_A}=M_{J_B}=1$
corresponds almost exclusively (99.98\%) to the quintet state.

\begin{figure}
\centering
\includegraphics[width=15cm]{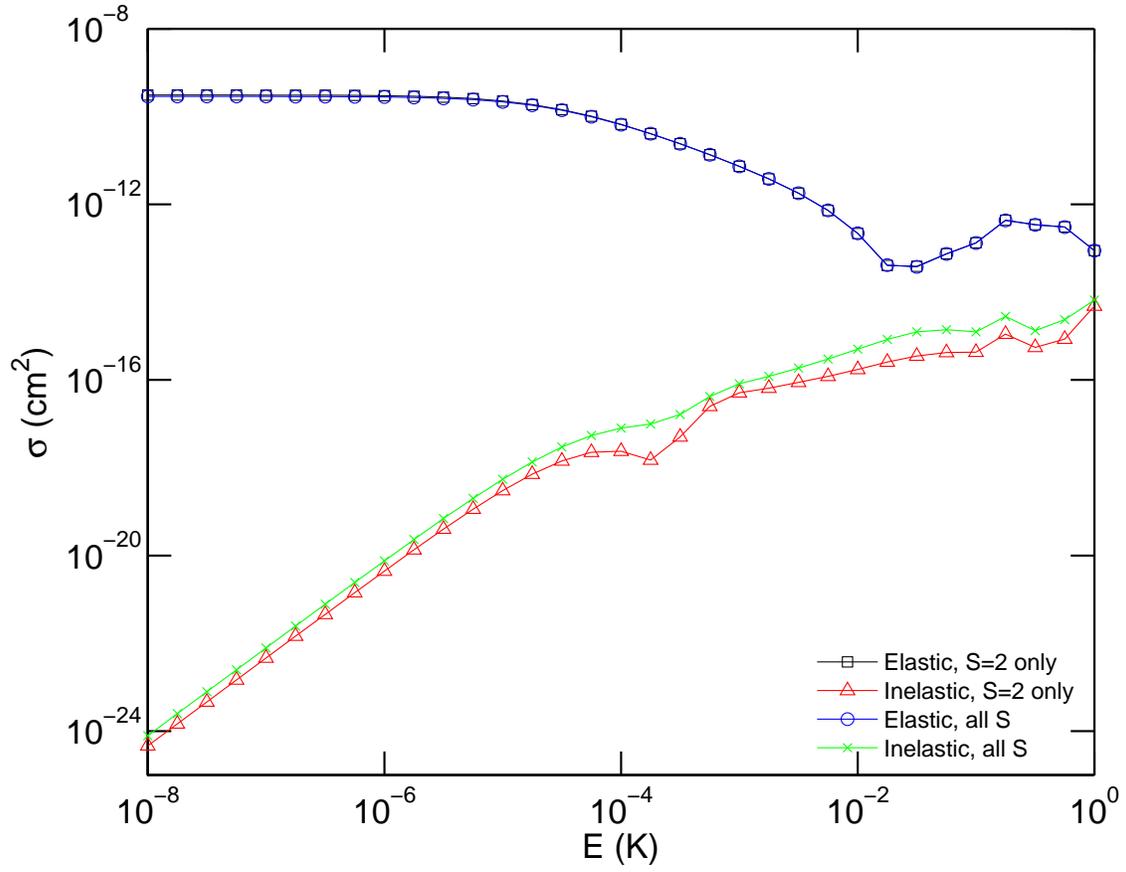} 
\caption
  {\label{fig:xsec_S=2only}
Elastic and inelastic $M_J$-changing cross sections for magnetically trapped $^{15}$NH
obtained from scattering calculations with only the quintet state included in the basis.
The cross sections calculated with all three spin states included (\textit{``all S"})
are shown for comparison.
}
\end{figure}

In order to investigate the contributions from the $S=0$ and 1 states, we have
performed scattering calculations with all singlet and triplet functions
removed from the basis set. The results are shown in Fig.\ \ref{fig:xsec_S=2only}
as a function of energy. The cross sections for the full basis set, i.e.\ with all three spin
states included, are also plotted for comparison. It can be seen that exclusion
of the $S=0$ and 1 states has a rather small effect on the cross section,
suggesting that most of the trap loss takes place within the quintet state.
Thus, the singlet and triplet states play a minor role in the collision
dynamics when described by the nonreactive $S=2$ potential.


If the $S=0$ and 1 states would be described by their true, reactive surfaces,
it can be expected that any transition to the singlet or triplet state leads to
chemical reaction and consequent trap loss.  In that case, however, the
potentials are no longer degenerate at short range and the probability for hopping from the
quintet surface to another state 
is most probably \textit{decreased} due to the energy
gap law. That is, inclusion of the reactive $S=0$ and 1 surfaces will probably
not lead to a larger inelastic cross section, and our assumption of including
only the nonreactive $S=2$ surface is very reasonable. In this respect, we may
also view the $M_J$-changing cross sections presented in Fig.\ \ref{fig:xsec_S=2only} as
approximate upper bounds. Nevertheless, it must be noted that the relatively
deep wells in the reactive potentials will give rise to a large number of bound
states, which in turn may cause strong resonances in the cross sections. 
In order to verify these assumptions, we plan to perform reactive quantum scattering
calculations for NH + NH with all three interaction potentials included.

\subsection{Sensitivity to potential and basis-set size}
\label{sec:results_C}
In this section we address two interrelated topics, namely the
sensitivity to the potential and the dependence on the angular basis-set size.
It is well established that low-energy scattering depends strongly on the
presence of bound and quasi-bound states near the dissociation threshold. 
Such states can give rise to scattering resonances that may enhance the collision
cross section by several orders of magnitude. The energies of these \mbox{(quasi-)}bound
states are highly sensitive to the details of the potential-energy surface, 
and hence they are very difficult to predict from first principles. 
Even a state-of-the art \textit{ab initio} potential cannot reliably predict
whether a particular near-dissociation state lies above or below the threshold.
This is particularly true for systems with multiple degrees of freedom and deep potential wells,
for which the density of states is relatively high. Thus, in order to assess the accuracy
of the cross sections, we must carefully take into account the effect of uncertainties in the potential.
In a related manner, we also consider the effect of using different channel
basis-set sizes in the scattering calculations. The size of the angular basis set can influence the
energies of the (quasi-)bound states, which in turn can lead to a different resonance structure.
It will be demonstrated, however, that the use of a reduced basis set leads only to a shift in the resonance
positions, and does not significantly alter the general resonance pattern. 





We first consider the sensitivity of the calculated cross sections to the potential-energy surface. 
Our potential has been obtained from state-of-the-art \textit{ab initio} calculations, and we
estimate that it differs from the exact potential by at most a few percent.
For practical reasons, we have studied the potential dependence indirectly
by performing scattering calculations as a function of the reduced mass $\mu$. Since scaling
the reduced mass by a factor of $\lambda$ ($\mu_{\rm{scaled}} = \lambda\mu$)
is almost equivalent to scaling the entire interaction potential by
$\lambda$ \cite{zuchowski:11}, this provides a
stringent test for the sensitivity to the potential.
The true potential does not necessarily differ from our \textit{ab initio} surface 
by only a constant factor,
but scaling by $\lambda$ ($0.9 \leq \lambda \leq 1.1$) amply samples the range of possibilities within which
the exact potential is expected to lie. 

\begin{figure}
\centering
\includegraphics[width=15cm]{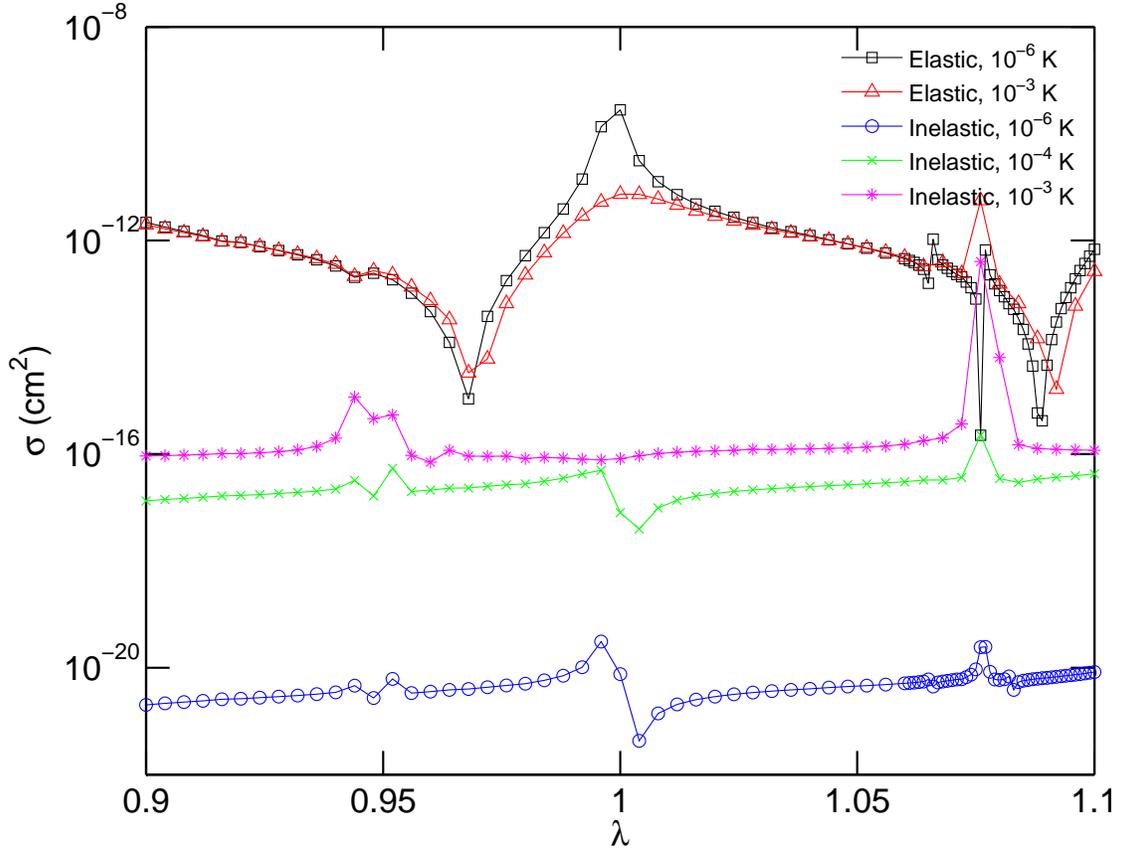} 
\caption
  {\label{fig:xsecmu}
Elastic and inelastic $M_J$-changing cross sections for magnetically trapped $^{15}$NH
as a function of the scaling parameter $\lambda$, calculated at collision energies of
$10^{-6}$ K, $10^{-4}$ K, and $10^{-3}$ K. The elastic cross sections for $10^{-4}$ K 
are the same as for $10^{-6}$ K. 
}
\end{figure}

Figure \ref{fig:xsecmu} shows the cross sections as a function of $\lambda$ at
collision energies of 10$^{-6}$ K, 10$^{-4}$ K, and 10$^{-3}$ K.
It can be seen 
that both the elastic and inelastic cross sections change by several orders
of magnitude as a function of $\lambda$, 
but they
vary about a certain background value. For instance, the elastic
cross sections fluctuate around $\sim10^{-12}$ cm$^2$ for all three collision energies.
The background values for the inelastic cross sections increase with $E^2$ in the ultracold regime,
consistent with the results of Fig.\ \ref{fig:xsec1415} and the
threshold laws discussed in Sec.\ \ref{sec:results_A}. 
The deviations from the background values 
are due to scattering resonances, which arise from  
NH--NH states that change from bound to
quasi-bound at the $| M_{J_A}=1, M_{J_B}=1 \rangle$ threshold.
Such resonance features are to be expected as a function of $\lambda$,
since a scaling of the potential, or in fact any modification of the potential-energy surface,
will cause a shift in the bound-state energies. 
For $10^{-6}$, $10^{-4}$ K, and $10^{-3}$ K, the resonances
are located around the same values of $\lambda$, and hence the $\lambda$-dependent resonance
structure would not be averaged out in a thermal (Maxwell-Boltzmann) distribution at temperatures
below 1 mK. 
That is, thermally averaged rate constants 
are likely to show a similar sensitivity to the potential as the calculated cross sections.

\begin{figure}
\centering
\includegraphics[width=15cm]{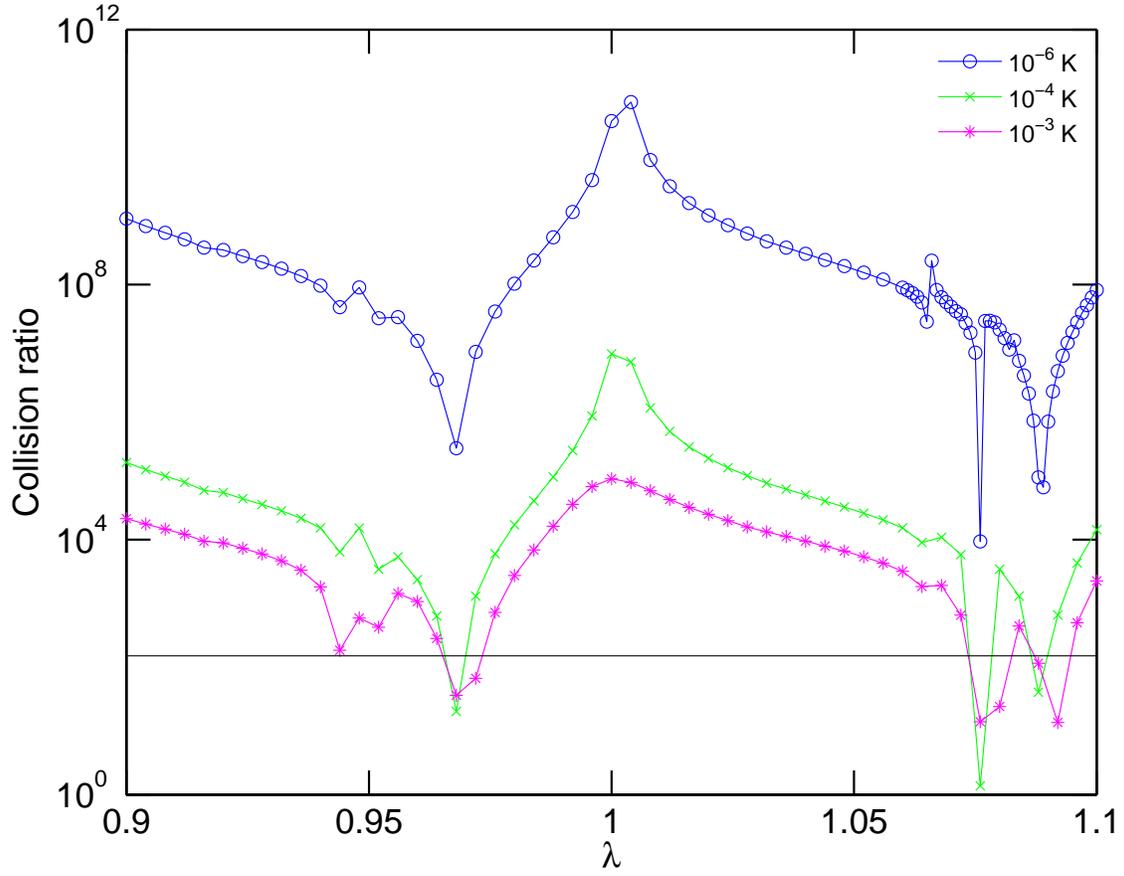} 
\caption
  {\label{fig:ratio_energies}
Elastic-to-inelastic cross-section ratios for magnetically trapped $^{15}$NH
as a function of the scaling parameter $\lambda$, calculated at collision energies of
$10^{-6}$ K, $10^{-4}$ K, and $10^{-3}$ K. The horizontal black line indicates the critical value of 150
that is required for efficient evaporative cooling.
}
\end{figure}

Let us now consider the elastic-to-inelastic cross-section ratios as a function of $\lambda$.
These are shown in Fig.\ \ref{fig:ratio_energies} for $E=10^{-6}$ K, 10$^{-4}$ K, and 10$^{-3}$ K.
For clarity, we have also indicated the critical ratio of 150 that is required for efficient
evaporative cooling. As can be seen, the calculated ratios exceed 150 for
almost all values of $\lambda$ and all energies considered, except when $\lambda$ is close to resonance.
This demonstrates that evaporative
cooling of NH is feasible at energies below 1 mK for most of the $\lambda$-values considered.
Although we cannot predict which value of $\lambda$ corresponds most closely 
to the exact potential, we do expect that the sampled range of $\lambda$ is indicative of the range within which the
exact potential lies, and hence we conclude that the \textit{probability} for successful evaporative cooling
is relatively large. That is, the true potential is very likely to be such that the elastic-to-inelastic ratio exceeds 150.

\begin{figure}
\centering
\includegraphics[width=15cm]{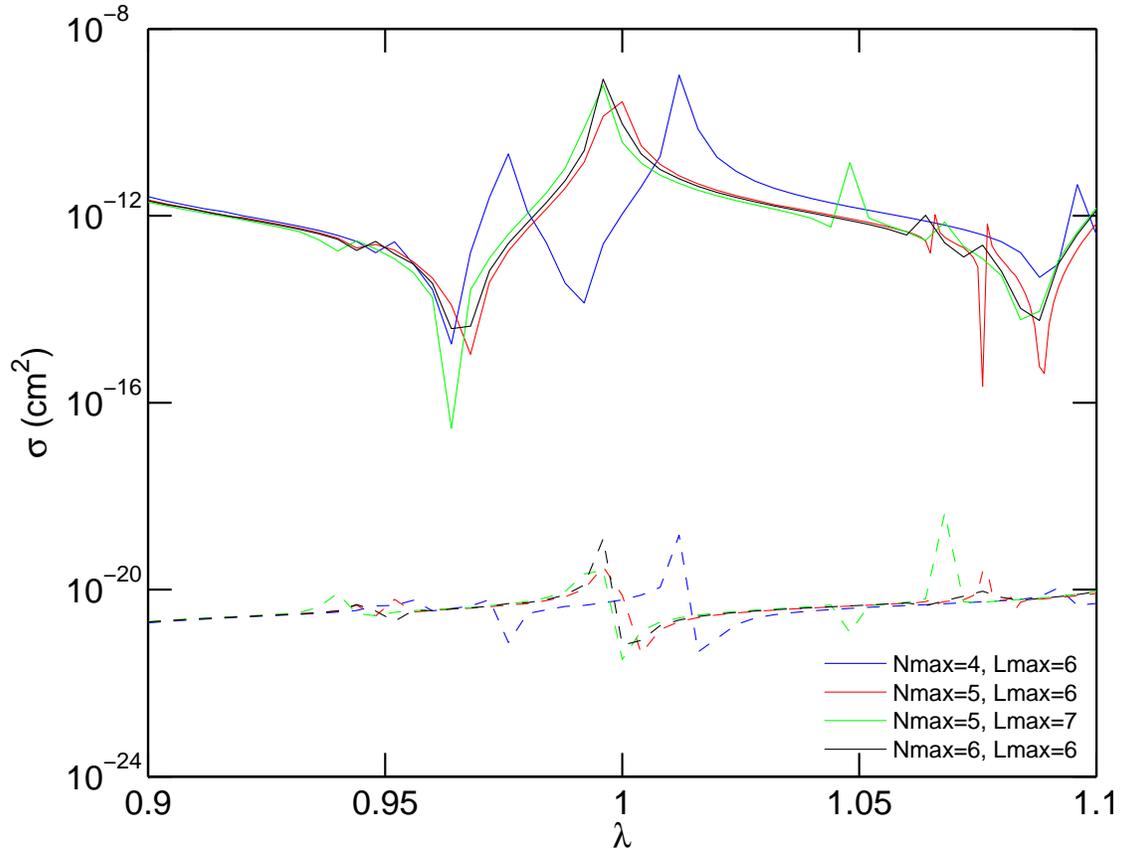} 
\caption
  {\label{fig:xsecmu_basis}
Elastic and inelastic $M_J$-changing cross sections for magnetically trapped $^{15}$NH
as a function of $\lambda$, calculated for different basis sets at a collision energy of $10^{-6}$ K.
Solid lines correspond to elastic cross sections and dashed lines to inelastic cross sections.
Different colors represent different basis sets.
}
\end{figure}

The $\lambda$-scaling approach is also used to investigate the influence of the angular basis-set
size on the scattering results. First we point out that the strong
anisotropy of the potential and the large reduced mass of NH--NH require
relatively high values of the basis-set parameters $N_{\rm{max}}$ and
$L_{\rm{max}}$. In addition, the triplet spins on the monomers increase the
channel basis-set size by a factor of 9, making it highly challenging to
achieve full basis-set convergence. Figure \ref{fig:xsecmu_basis} shows the
cross sections as a function of $\lambda$ for different values of
$N_{\rm{max}}$ and $L_{\rm{max}}$ at a collision energy of 10$^{-6}$ K.
The maximum number of channels in these
calculations ranged from 937 for $N_{\rm{max}}=4$ and $L_{\rm{max}}=6$
($\mathcal{J}=4$) up to 2382 for $N_{\rm{max}}=6$ and $L_{\rm{max}}=6$
($\mathcal{J}=5$). 
It can be seen that the cross sections 
all vary by several orders of magnitude as a function of $\lambda$,
and for a given value of $\lambda$ the four basis sets can
yield very different numerical results. 
However, the different cross sections 
vary about the same background values and the resonant
features have similar widths for all four basis sets.
Thus, a change in $N_{\rm{max}}$ or
$L_{\rm{max}}$ may cause a shift in the positions of the resonances,
but the overall pattern is virtually unaffected.
The estimated probability for successful evaporative cooling, i.e.\
the probability that the exact potential is such that the elastic-to-inelastic
cross-section ratio exceeds 150, is therefore similar for all four basis sets.
This can also be understood by considering that a change
in the basis set only shifts the bound-state energy levels,
similar to the effect of scaling the potential.

The results of Fig.\ \ref{fig:xsecmu_basis} demonstrate that the cross sections are
almost, but not fully converged with respect to $N_{\rm{max}}$ and $L_{\rm{max}}$. Using a larger
basis set is infeasible at present given the available computer power.
A larger basis set would also require additional terms in the expansion of
the potential anisotropy [Eq.\ (\ref{eq:V_sf})], making the calculation prohibitively
expensive. 
Moreover,
taking into account the uncertainty in the potential, even a fully converged
basis set would not give really reliable numerical values due to the presence
of (quasi-)bound state resonances. Since the exact form of the potential, and
thus the precise locations of the resonances, are still unknown, the calculated
cross sections are subject to an inherent degree of uncertainty that cannot be
reduced by the use of a fully converged basis set.
In this sense, full basis-set convergence will not necessarily yield a more
accurate prediction of the true cross sections.
On the other hand, 
the \textit{probability} for successful evaporative cooling can be reliably
predicted using an incompletely converged basis set, and hence we conclude 
that, even if full basis-set convergence could be achieved, this would not
significantly alter our main qualitative results.
We emphasize, however, that it is crucial to test the sensitivity to the potential
in order to assess the accuracy of the calculated cross sections.
As a final point, we note that the
uncertainty limits of the potential could, in principle, be greatly reduced by
measuring the cross sections experimentally.

\section{Conclusions}
\label{sec:concl}
We have carried out elastic and inelastic quantum scattering calculations on a state-of-the-art
\textit{ab initio} potential to
study field-free NH + NH collisions at low and ultralow temperatures. 
The results indicate that, when the molecules are prepared in their nuclear
spin-stretched states, bosonic $^{15}$NH is more suitable for evaporative
cooling than fermionic $^{14}$NH. This is a direct consequence of the Pauli principle,
which forbids $s$-wave scattering for two identical fermions.
The $^{14}$NH isotope may also be successfully cooled,
however, when the monomers are in a mixture of different nuclear spin states.

We have assumed that all three spin states of the NH--NH complex are
described by the nonreactive quintet surface. This approximation is shown to be
reasonable, although a full reactive scattering calculation would be
required to investigate the precise role of the chemically active singlet and
triplet states.

The collision cross sections are sensitive to the details of
the interaction potential, because of the presence of quasi-bound states that
cause scattering resonances. Since the exact interaction potential is unknown,
this gives rise to a degree of uncertainty in the numerical cross sections.
However, a sampling of the range of possibilities indicates
that the exact
potential is very likely to be such that the elastic-to-inelastic cross-section
ratio is favorable for evaporative cooling.
This result is only weakly dependent on
the size of the channel basis set.
In particular, 
the effect of using a reduced basis set is very similar
to a scaling of the potential within its uncertainty. 
We conclude that even without full basis-set convergence, which
is extremely difficult to achieve for systems such as NH--NH, we can 
provide valuable insight into the feasibility of evaporative cooling.
This also offers hope for the theoretical treatment of
other challenging open-shell molecule + molecule systems.

\begin{acknowledgments}
We gratefully acknowledge EPSRC for funding the collaborative project CoPoMol
under the ESF EUROCORES programme EuroQUAM. LMCJ and GCG thank the Council for
Chemical Sciences of the Netherlands Organization for Scientific Research
(CW-NWO) for financial support.
\end{acknowledgments}

\appendix*
\section{Matrix elements}
\label{sec:appendix}
In this Appendix, we present
the matrix elements of the scattering Hamiltonian in the
`primitive' basis $\psi^{\mathcal{J},\mathcal{M}}_{N_A,N_B,N,S_A,S_B,S,J,L}$.
The matrix elements in the symmetry-adapted basis can be obtained
using Eq.\ (\ref{eq:basis_sa}).
For the angular functions of the potential we have
\begin{eqnarray}
\label{eq:ALLL_ij}
\lefteqn{\langle \psi^{\mathcal{J},\mathcal{M}}_{N_A,N_B,N,S_A,S_B,S,J,L} |
A_{L_A,L_B,L_{AB}} |
\psi^{\mathcal{J},\mathcal{M}}_{N_A',N_B',N',S_A,S_B,S',J',L'} \rangle = } \nonumber \\
& & \hphantom{X}\delta_{SS'} \left(\frac{1}{4\pi}\right)^{3/2} (-1)^{N_A+N_B+N+S+L_{AB}+\mathcal{J}} \nonumber \\
& & \hphantom{X}\times [L_{AB}] \sqrt{[L_A][L_B][N_A][N_A'][N_B][N_B'][N][N'][L][L'][J][J']} \hphantom{XX} \nonumber \\
& & \hphantom{X}\times\left( \begin{array}{ccc} N_A & L_A & N_A' \\ 0 & 0 & 0 \end{array} \right)
  \left( \begin{array}{ccc} N_B & L_B & N_B' \\ 0 & 0 & 0 \end{array} \right)
  \left( \begin{array}{ccc} L & L_{AB} & L' \\ 0 & 0 & 0 \end{array} \right) \nonumber \\
& & \hphantom{X}\times \left\{ \begin{array}{ccc} J & J' & L_{AB} \\ L' & L & \mathcal{J} \end{array} \right\}
  \left\{ \begin{array}{ccc} N' & N & L_{AB} \\ J & J' & S \end{array} \right\}
  \left\{ \begin{array}{ccc} N_A & N_A' & L_A \\ N_B & N_B' & L_B \\ N & N' & L_{AB} \end{array} \right\},
\end{eqnarray}
with the factors in large round brackets denoting Wigner $3j$ symbols, the factors in curly brackets
denoting $6j$ and $9j$ symbols, and $[Q] = (2Q+1)$.
The intermolecular magnetic dipole term is given by:
\begin{eqnarray}
\label{eq:Vmagndip_ij}
\lefteqn{\langle \psi^{\mathcal{J},\mathcal{M}}_{N_A,N_B,N,S_A,S_B,S,J,L} |
V_{\rm{magn.dip}} |
\psi^{\mathcal{J},\mathcal{M}}_{N_A',N_B',N',S_A,S_B,S',J',L'} \rangle =} \nonumber \\
& &\hphantom{X} -\delta_{N_A N_A'} \delta_{N_B N_B'} \delta_{NN'} \sqrt{30} g_S^2 \mu_{\rm{B}}^2 \frac{\alpha^2}{R^3}
(-1)^{N+S'+J+J'+\mathcal{J}} \nonumber \\
& &\hphantom{X} \times \sqrt{ S_A(S_A+1)S_B(S_B+1)[S_A][S_B][S][S'][J][J'][L][L'] } \nonumber \\
& &\hphantom{X} \times \left( \begin{array}{ccc} L & 2 & L' \\ 0 & 0 & 0 \end{array} \right)
\left\{ \begin{array}{ccc} J & J' & 2 \\ L' & L & \mathcal{J} \end{array} \right\}
\left\{ \begin{array}{ccc} J' & J & 2 \\ S & S' & N \end{array} \right\}
\left\{ \begin{array}{ccc} S_A & S_A & 1 \\ S_B & S_B & 1 \\ S & S' & 2 \end{array} \right\}. \hphantom{XX}
\end{eqnarray}
The rotation operators for the two monomers ($i=A,B$) are completely diagonal in the angular basis:
\begin{eqnarray}
\label{eq:Hrot_ij}
\lefteqn{\langle \psi^{\mathcal{J},\mathcal{M}}_{N_A,N_B,N,S_A,S_B,S,J,L} |
B_0 \hat{N}^2_i |
\psi^{\mathcal{J},\mathcal{M}}_{N_A',N_B',N',S_A,S_B,S',J',L'} \rangle =} \nonumber \\
& & \hphantom{X}\delta_{N_A N_A'}\delta_{N_B N_B'}\delta_{N N'}\delta_{S S'}\delta_{J J'}\delta_{L L'} B_0 N_i(N_i+1).
\hphantom{XXX}
\end{eqnarray}
For the spin-rotation coupling terms we find
\begin{eqnarray}
\label{eq:HspinrotA_ij}
\lefteqn{\langle \psi^{\mathcal{J},\mathcal{M}}_{N_A,N_B,N,S_A,S_B,S,J,L} |
\gamma \hat{\bm{N}}_A\cdot\hat{\bm{S}}_A |
\psi^{\mathcal{J},\mathcal{M}}_{N_A',N_B',N',S_A,S_B,S',J',L'} \rangle =} \nonumber \\
& &\hphantom{X} \delta_{N_A N_A'}\delta_{N_B N_B'}\delta_{J J'}\delta_{L L'} \gamma
(-1)^{N_A+N_B+S_A+S_B+S+S'+J} \nonumber \\
& &\hphantom{X} \times \sqrt{N_A(N_A+1)S_A(S_A+1)[N_A][S_A][N][N'][S][S']} \nonumber \\
& &\hphantom{X} \times \left\{ \begin{array}{ccc} N_A & N_A & 1 \\ N & N' & N_B \end{array} \right\}
\left\{ \begin{array}{ccc} S_A & S_A & 1 \\ S & S' & S_B \end{array} \right\}
\left\{ \begin{array}{ccc} N & N' & 1 \\ S' & S & J \end{array} \right\},
\hphantom{X}
\end{eqnarray}
\begin{eqnarray}
\label{eq:HspinrotB_ij}
\lefteqn{\langle \psi^{\mathcal{J},\mathcal{M}}_{N_A,N_B,N,S_A,S_B,S,J,L} |
\gamma \hat{\bm{N}}_B\cdot\hat{\bm{S}}_B |
\psi^{\mathcal{J},\mathcal{M}}_{N_A',N_B',N',S_A,S_B,S',J',L'} \rangle =} \nonumber \\
& & \hphantom{X}\delta_{N_A N_A'}\delta_{N_B N_B'}\delta_{J J'}\delta_{L L'} \gamma
(-1)^{N_A+N_B+N+N'S_A+S_B+J} \nonumber \\
& & \hphantom{X} \times \sqrt{N_B(N_B+1)S_B(S_B+1)[N_B][S_B][N][N'][S][S']} \nonumber \\
& & \hphantom{X} \times \left\{ \begin{array}{ccc} N' & N & 1 \\ N_B & N_B & N_A \end{array} \right\}
\left\{ \begin{array}{ccc} S' & S & 1 \\ S_B & S_B & S_A \end{array} \right\}
\left\{ \begin{array}{ccc} N & N' & 1 \\ S' & S & J \end{array} \right\},
\hphantom{X}
\end{eqnarray}
and, finally, for the intramolecular spin-spin operators $\hat{V}^{(i)}_{\rm{SS}}$ we have
\begin{eqnarray}
\label{eq:HspspA_ij}
\lefteqn{ \langle \psi^{\mathcal{J},\mathcal{M}}_{N_A,N_B,N,S_A,S_B,S,J,L} |
\hat{V}^{(A)}_{\rm{SS}} |
\psi^{\mathcal{J},\mathcal{M}}_{N_A',N_B',N',S_A,S_B,S',J',L'} \rangle = } \nonumber \\
& & \hphantom{X} \delta_{N_B N_B'}\delta_{J J'}\delta_{L L'} \frac{2}{3} \sqrt{30} \lambda_{\rm{SS}}
(-1)^{N_B+S_A+S_B+S+S'+J} \nonumber \\
& & \hphantom{X}\times S_A(S_A+1)(2S_A+1) \sqrt{[N_A][N_A'][N][N'][S][S']}
\left( \begin{array}{ccc} N_A & 2 & N_A' \\ 0 & 0 & 0 \end{array} \right) \nonumber \\
& & \hphantom{X}\times \left\{ \begin{array}{ccc} N_A' & N_A & 2 \\ N & N' & N_B \end{array} \right\}
\left\{ \begin{array}{ccc} S_A & S_A & 1 \\ 1 & 2 & S_A \end{array} \right\}
\left\{ \begin{array}{ccc} S_A & S_A & 2 \\ S & S' & S_B \end{array} \right\}
\left\{ \begin{array}{ccc} N & N' & 2 \\ S' & S & J \end{array} \right\},
\hphantom{X}
\end{eqnarray}
\begin{eqnarray}
\label{eq:HspspB_ij}
\lefteqn{ \langle \psi^{\mathcal{J},\mathcal{M}}_{N_A,N_B,N,S_A,S_B,S,J,L} |
\hat{V}^{(B)}_{\rm{SS}} |
\psi^{\mathcal{J},\mathcal{M}}_{N_A',N_B',N',S_A,S_B,S',J',L'} \rangle = } \nonumber \\
& & \hphantom{X} \delta_{N_A N_A'}\delta_{J J'}\delta_{L L'} \frac{2}{3} \sqrt{30} \lambda_{\rm{SS}}
(-1)^{N_A+N_B+N_B'+N+N'+S_A+S_B+J} \nonumber \\
& & \hphantom{X}\times S_B(S_B+1)(2S_B+1) \sqrt{[N_B][N_B'][N][N'][S][S']}
\left( \begin{array}{ccc} N_B & 2 & N_B' \\ 0 & 0 & 0 \end{array} \right) \nonumber \\
& & \hphantom{X}\times \left\{ \begin{array}{ccc} N' & N & 2 \\ N_B & N_B' & N_A \end{array} \right\}
\left\{ \begin{array}{ccc} S_B & S_B & 1 \\ 1 & 2 & S_B \end{array} \right\}
\left\{ \begin{array}{ccc} S' & S & 2 \\ S_B & S_B & S_A \end{array} \right\}
\left\{ \begin{array}{ccc} N & N' & 2 \\ S' & S & J \end{array} \right\}.
\hphantom{X}
\end{eqnarray}

\bibliography{vanderwaals}

\clearpage

\clearpage

\clearpage

\clearpage

\clearpage

\clearpage

\clearpage


\end{document}